\documentclass[aps,eqsecnum,12pt,amsfonts]{revtex4}
\usepackage{amsfonts}
\usepackage{amsmath}
\usepackage{bm}

\newcommand{\ket}[1]{| #1    \rangle }
\newcommand{\bra}[1]{ \langle   #1  | }

\newcommand{\mel}[3]{  \langle #1  | #2   | #3  \rangle }
\newcommand{\amp }[2]{ \langle #1 |  #2  \rangle }

\newcommand{\tr}[1]{{\rm Tr  }(  #1  ) }
\newcommand{\pal}{\parallel}

\newcommand{\D}{{\form{ d}}}
\newcommand{\DD}{{\form{D}}}
\newcommand{\psif}{\psi_f}
\newcommand{\psii}{\psi_i}

\newcommand{\mels}[1]{\mel{\psif}{#1}{\psii}}

\newcommand{\phs}{\eta}
\newcommand{\hatu}{\hat{U}}

\newcommand{\gl}{\hat{\lambda}}
\newcommand{\form}[1]{{\bm{  #1}}}
\newcommand{\vect}[1]{{\bm{ #1}}}
\newcommand{\fre}{\vect{e}}
\newcommand{\from}{\form{\omega}}

\newcommand{\sub}[1]{_{{}_{#1}}}

\newcommand{\rref}[1]{~(\ref{#1})}
\newcommand{\ccite}[1]{~\cite{#1}}

\begin{document}


\title[Phase/Modulo Relations]{ Geometric phase and modulus relations for  SU(n) matrix elements in the defining representation}

\author{ Alonso Botero }
\email{abotero@uniandes.edu.co} 
\affiliation{
    Departamento de F\'{\i}sica,
    Universidad de los Andes,
    Apartado A\'ereo 4976, Bogot\'a, Colombia}

\date{\today}

\begin{abstract}

A set of relations between the modulus and phase is derived for amplitudes of the form $\mels{\hatu(x)}$ where $\hat{U}(x) \in SU(n)$ in the fundamental representation and $x$ denotes the coordinates on the group manifold. An illustration is given for the case $n=2$ as well as a brief discussion of phase singularities and superoscillatory phase behavior for such amplitudes. The present results complement results obtained previously~\cite{PMrel1}  for amplitudes valued on the ray space ${\cal R} = {\mathbb C}P^n$. The connection between the two is discussed. 

\end{abstract}
\maketitle

\pagebreak

\section{Introduction}

In a previous paper,~\cite{PMrel1} a number of relations have been
obtained for the interdependence between the phase and modulo of
 amplitudes of the form $\amp{\psi_f}{\psi(x)}$, where
$\ket{\psi_f}$ is some fixed state and $\ket{\psi(x)}$ is
parameterized on a complex parameter subspace $\mathcal M$ of the
ray space ${\mathcal R}$, with general curvilinear real coordinates
$\{x\}$. In particular, it has been shown that the phase $\eta(x)
= \arg \amp{\psi_f}{\psi(x)}$ and modulus
$\sqrt{p(x)}=|\amp{\psi_f}{\psi}|$ are intimately connected
functions on $\mathcal M$ through the relation
\begin{equation}
\label{cauchgen}
\nabla \log \sqrt{p}  = \ \ \form{\Omega} \left( \nabla \eta - \vect{A} \right) 
\end{equation}
where $\form{\Omega}$ is the K\"ahler  2-form  on $\cal M$ and
$\form{A}$ is the Berry-Simon \ccite{berry,simon} connection  $\vect{A} = - i
\amp{\psi(x)}{\nabla \psi(x)}$. Such conditions constitute a
generalization of ordinary Cauchy-Riemann conditions, and reflect
the locally holomorphic nature of such spaces. A number of
interesting results follow when $\ket{\psi(x)}$  sweeps the
whole space of quantum states, i.e., the full ray space ${\cal
R}$. Using relations \rref{cauchgen} and the
relation  between transition probabilities
and geodesic distances as measured with the Fubini-Study metric\ccite{anaha90b},
it then becomes possible to show  that
\begin{subequations}\label{cprel}
\begin{eqnarray}
q |\nabla \eta -\form{A}|^2 & = & \frac{1}{p} -1\label{cprela}\\
 q |\nabla \log \sqrt{p}|^2 & = & \frac{1}{p} -1  \label{cprelb}\\
\nabla p\cdot(\nabla \eta - \form{A}) &= &0 \label{cprelc}\, ,
\end{eqnarray}
\end{subequations}
where $q$ is an arbitrary overall scale parameter in the
definition of the  metric.  In particular, it follows
from \rref{cprela} that
\begin{equation}
\label{ampcp}
 \amp{\psif}{\psi(x)}= \frac{ e^{i \phs(x)} }{ \sqrt{1 +
q| \nabla \eta - \form{A} |^2}} \, ,
\end{equation}
implying that the transition amplitude can be expressed entirely
in terms of its phase dependence.

In this paper we present a similar set of relations for amplitudes
of the form $\mels{U(x)}$ where $\hatu(x)$ is an element of
$SU(n)$ in the defining ($n$-dimensional) representation, $x$ are
now coordinates coordinates on the group manifold, and
$\ket{\psi_i}$ and $\ket{\psi_f}$ are any two fixed normalized
vectors acting on an $n$-dimensional Hilbert space:

 Let $p(x)$ and
$\eta(x)$ be defined respectively as the modulus squared and phase
angle of $\mels{\hat{U}(x)}$. It then becomes possible to show
that
\begin{subequations}
\label{surel}
\begin{eqnarray}
|\nabla \phs|^2  & = & \frac{1}{p} + \left( 1 - \frac{2}{n} \right)\, , \label{surela}\\
|\nabla \log \sqrt p|^2   & = &  \frac{1}{p} - 1 , \label{surelb}\\
\nabla p \cdot \nabla \eta & = & 0 \label{surelc}\, ,
\end{eqnarray}
\end{subequations}
where the inner product is now taken with respect to the
Cartan-Killing metric on the group manifold, expressible as
\begin{equation}
g_{\mu \nu}= \frac{1}{2}{\rm Tr}( \partial_\mu \hatu \ \partial_\nu \hatu ^\dag ) \, .
\end{equation}
In particular,
\rref{surela} implies that the amplitude can be parameterized
entirely in terms of its phase according to
\begin{equation}
\label{ampsun} \mels{\hat{U}(x)}         = \frac{ e^{i\phs(x)} }{
\sqrt{ | \nabla \eta |^2  - ( n-2)/n } }\, ,
\end{equation}
in a similar fashion to \rref{ampcp}.

That there should exist a connection between relations
\rref{cprel} and \rref{surel} may be inferred from the fact that
$SU(n)$ is a principal $U(n-1)$-bundle  over the coset
space $SU(n)/U(n-1)$, in which the $U(n-1)$ corresponds to the
isotropy group leaving a fixed ray in Hilbert space invariant. 
Relations \rref{cprel} may then be viewed as the
set of gauge-invariant relations obtained from \rref{surel} after
``modding out" the subgroup of $SU(n)$  not affecting the angle
between rays, in other words, the transition probability.
Relations \rref{surel} will be proved in the next section and in
the final section the connection between \rref{cprel} and
\rref{surel} will be established.

Before proceeding with the proofs, however, an illustration
of relation \rref{ampsun} and some of its consequences may be useful. In the case of $SU(2)$, \rref{ampsun} takes the particularly simple form
\begin{equation}\label{ampsun2}
\mels{\hat{U}(x)}         = \frac{ e^{i\phs(x)} }{
 | \nabla \eta | } \, .
\end{equation}
Thus, let the initial and final states be $ | i \rangle = |+ \rangle $ and $|
f \rangle = |- \rangle $ in standard spin-$1/2$ notation. Since
the group manifold for $SU(2)$ is $S^3$, it becomes convenient to
introduce a standard polar coordinate chart on the three-sphere
$x=$($\chi$, $\theta$, $\phi$), where $\chi,\ \theta \in (0,\pi)$
and $\phi \in [0,2 \pi)$. In terms of these coordinates, a natural
parameterization of an $SU(2)$ group element is then
\begin{equation}
\hat{U}(x) = \cos\!\chi \openone + i\, \sin\chi\ \vec{\sigma}\cdot
\hat{n}(\theta, \phi)\, ,
\end{equation}
with a corresponding metric element on $S^3$
\begin{equation}
g_{\mu \nu} dx^\mu dx^\nu = d\chi^2 + \sin^2\!\chi ( d\theta^2 +
\sin^2\!\theta \ d\phi^2 ) \, .
\end{equation}
Note that $\hat{U}(x) $ corresponds, in the language of angular
momentum, to a spacial rotation by an angle $2 \chi $ around the
axis $\hat{n}(\theta, \phi)$ (the unit vector on the two-sphere).
A simple calculation then shows that
\begin{equation}
\langle -| \hat{U}(x) | + \rangle = i \sin \chi \sin \theta e^{i
\phi} \, ,
\end{equation}
from which we  identify
\begin{equation}
\phs(x) = \phi + \pi/2 \, , \ \ \  p(x) = \sin^2\! \chi \sin^2\!
\theta \, .
\end{equation}
Noting  that the  inverse metric components are $g^{\chi \chi} =
1$, $ g^{\theta \theta} = 1/\sin^2{\chi}$,  $ g^{\phi \phi} = 1
/\sin^2{\chi} \sin^2{\theta}$, and all others vanishing, we
therefore see that
\begin{equation}
|\nabla \phs |^2 = g^{\phi \phi} (\partial_\phi \phs)^2=
\frac{1}{\sin^2\! \chi \sin^2\!  \theta } \,  = 1/p(x)\, ,
\end{equation}
in consistency with Eq. \rref{ampsun2}.

Two interesting consequences also emerge from equation \rref{ampsun}: First, we see that as $x$ approaches a value $x_o$ such that  $\mels{\hatu(x_o)}=0$, the phase gradient must diverge. The previous illustration shows that in the case of $SU(2)$, this divergence shows  vortex  behavior about a  string singularity (for the above initial and final states the singular string corresponds to the $S^3$ meridians at $\theta = 0$ and $\pi$  running from $\chi=0$ to $\chi=\pi$). 

The other interesting consequence is that since the modulus may never exceed unity, the phase gradient is  bounded from below and is therefore never allowed to vanish. In other words, the phase factor cannot be stationary on the $SU(n)$ manifold. It is amusing to note that the lower bound on the gradient
\begin{equation}
|\nabla \phs |_{min} = \sqrt{\frac{2(n-1)}{n}}
\end{equation}
in fact corresponds to an {\em upper bound} on the magnitude of the eigenvalues of any generator $\hat{l}$ of $SU(n)$ normalized such that $Tr(\hat{l}^2) = 2$. To see the implications of this, note that a single parameter transformation $\hatu(t) = \exp(i \hat{l}\,t  )$ generates a curve on $SU(n)$ in which the curve length with respect to the Cartan-Killing metric is $ds = dt$. If  $\hat{l}$ is now chosen in the direction of the phase gradient evaluated at the identity ($\hatu = \openone$), then  the local angular frequency of the phase oscillation  $\omega(t)$, defined as
\begin{equation}
\omega(t) = \frac{d}{dt}\arg \mels{\exp(i \hat{l}\,t )} 
\end{equation}
corresponds, at $t = 0$ to the phase gradient $|\nabla \phs|$ evaluated at the identity element, and must therefore satisfy  $\omega(0) \geq \sqrt{\frac{2(n-1)}{n}}$. On the other hand,  $\mels{\exp(i \hat{l}\,t )} $ has a Fourier expansion of the form \begin{equation}
\mels{\exp(i \hat{l}\,t)} = \sum_k C_k e^{i  l_k\, t }
\end{equation}
where $l_k$ are the eigenvalues of $\hat{l}$, none of which may exceed in magnitude the value $\sqrt{\frac{2(n-1)}{n}}$. Thus, for any two given states $\ket{\psi_i}$ and $\ket{\psi_f}$ there always exists a generator $\hat{l}$ such that around $t=0$ the function $\mels{\exp(i t \hat{l} )}$ exhibits so-called super-oscillatory behavior\cite{BerrySuper,Kempf}: local phase oscillations which are at least as fast as those of the fastest Fourier component.

\section{Proof of Relations (\ref{surel})}
\label{relations}
\bigskip

Let $\{ \gl_a | a = 1, .. n^2 -1 \}$ be a set of
linearly-independent, traceless matrix generators for $SU(n)$ in
the fundamental ($n$-dimensional) representation, chosen so that
they satisfy the matrix inner product $\tr{\gl_a \gl_b } = 2
\delta_{ab}$. A euclidean inner product is  naturally
induced on the Lie algebra, with the metric form
\begin{equation}
\eta_{ab} = \frac{1}{2}\tr{ \gl_a \gl_b } \ \ \ (\ = \delta_{ab}\
) \, ,
\end{equation}
coinciding with the so-called Cartan-Killing form ~\cite{kobayashi}. Now consider  an open covering
of $SU(n)$,  parameterized by the matrix $\hat{U}(x) \in SU(n)$,
where $x$ stands for $n^2 -1$ coordinates
$\{x^\mu:\mu=1,..,n^2-1\}$.  A set of left-invariant one-forms $\{
\from^{a} \}$ is defined by the expansion of  the Lie-algebra
valued 1-form $\hatu^\dag \D\hatu  $  as a   linear
combination of the $\gl$-matrices
\begin{equation}\label{leftin}
\hatu^\dag \D\hatu = i \from^{a}\, \gl_{a} \, .
\end{equation}
An invariant metric tensor on the group manifold is then naturally
inherited from the Cartan-Killing according to
\begin{equation}
g = \eta_{ab}\, \from^{a} \otimes \from^{b}  = \frac{1}{2} \tr{
\D\hatu \otimes \D\hatu^\dag} \, .
\end{equation}
where the left-invariant forms play the role of a {\em vielbein}.
Similarly, the inverse metric tensor
 \begin{equation}
g^{-1} = \eta^{ab}\, \fre_{a} \otimes \fre_{b}
\end{equation}
is defined from the set of  vector fields $\{ \fre_a \}$ dual
to the left invariant forms, i.e., satisfying $
\from^{a}(\fre_{b}) = \delta^{a}_{b} \, $.

Now turn  to matrix elements of the form $\mels{\hatu }$. For
simplicity, let us represent this quantity  either in terms of its
two real polar components $\sqrt{p}$ and $\phs$, or in terms of a
complex phase $\chi$:
\begin{equation}
\mels{\hatu } = \sqrt{p}\, e^{i \phs} = e^{i \chi}
\end{equation}
Using \rref{leftin}, it is then easy to show that
\begin{equation}
\D \chi = \D\phs  - i \D \log \sqrt{p} = \from^{a} \frac{\mels{\hatu
\gl_a} }{\mels{\hatu}}\, .
\end{equation}
Thus, using the definition of the Cartan-Killing metric we can
then show that $\nabla\chi \cdot \nabla\chi^*=g^{-1}(\D \chi ,
\D\chi^*)$  and $\nabla\chi \cdot \nabla\chi^*=g^{-1}(\D \chi ,
\D\chi)$ can be expressed as
\begin{eqnarray}
\nabla\chi \cdot \nabla\chi^*   & = &
        \eta^{a b} \frac{\mels{\hatu
\gl_a} }{\mels{\hatu}}\frac{\mel{\psii}{\gl_b \hatu^\dag}{\psif} }{\mel{\psii}{\gl_b \hatu^\dag}{\psif}}  \nonumber\\
\nabla\chi \cdot \nabla\chi & = &  \eta^{a b} \frac{\mels{\hatu
\gl_a} }{\mels{\hatu}}\frac{\mels{\hatu
\gl_b} }{\mels{\hatu}} \, .
\end{eqnarray}
The right-hand sides of these two equations can be computed by
using the following identity on the fundamental representation of
$SU(n)$ (See e.g., ~\cite{greinmull})
\begin{equation}
\frac{1}{2}\eta^{ab}\tr{\hat{X}\, \gl_a\,  \hat{Y}\,  \gl_b } =
\tr{\hat{X}} \tr{\hat{Y}} - \frac{1}{n} \tr{\hat{X} \hat{Y} } \, ,
\end{equation}
where for the first one we use $\hat{X} =
\hatu^\dag\ket{\psif}\bra{\psif}\hatu $ and $\hat{Y}  =
\ket{\psii}\bra{\psii} $, and for the second one $\hat{X} =
\hat{Y} = \ket{\psii}\bra{\psif}\hatu$. Thus we find that
\begin{eqnarray}
\nabla\chi \cdot \nabla\chi^* & = &  2\left(\frac{1}{p} - \frac{1}{n} \right)    \nonumber\\
\nabla\chi \cdot \nabla\chi & = & 2\left(1 - \frac{1}{n} \right)
\, .
\end{eqnarray}
Re-expressing the gradient 1-form $d\chi$ in terms of $\phs$ and
$p$, and taking real and imaginary parts, one  obtains
(\ref{surela}-\ref{surelc}).

\section{Connection between relations (\ref{surel}) and  relations (\ref{cprel})}

To connect \rref{surel} and \rref{cprel}, we implement the
so-called Cartan decomposition of the Lie algebra ${\cal
L}(SU(n))$~\cite{kobayashi}. Let $\{\gl_i \}$, with  $i$ ranging from $0$ to
$(n-1)^2 -1 = n^2-2n$ span the Lie algebra of an isotropy group
$U(n-1)=SU(n-1)\times U(1) $, a subgroup of $SU(n)$ in which the
$SU(n-1)$ generated by $\gl_{i} $ acts on the
orthogonal subspace to $\ket{\psi_1}$ and the $U(1)$  generated by
$\gl_0$ implements a phase transformation on $|\psi_1\rangle$
and commutes with the $SU(n-1)$. The remaining $n^2 -1 - (n^2 - 2n
+1) = 2(n-1)$ generators of $SU(n)$, denoted by $\{ \gl_A \}$,
span an orthogonal complement in ${\cal L}(SU(n))$ to the Lie
algebra of the $U(n-1)$, in the sense of the Cartan-Killing form (CK).
For our purposes, it suffices to give  $\gl_0$ and the $\gl_A$
explicitly. Given the tracelessness condition plus the
normalization condition ${\rm Tr}(\gl_o^2 )=2$, the form of
$\gl_o$ is determined up to a sign, and we choose
\begin{equation}
\gl_o = \sqrt{\frac{2}{n(n-1)}}
 \sum_{k=1}^{n-1}\openone  - \sqrt{\frac{2 n }{n-1}}|\psii \rangle \langle \psii | \, .
\end{equation}
For the orthogonal generators $\{\hat{\lambda}_A \}$ we choose
matrices of the form
 \begin{equation}
\hat{X}_k = | \psii \rangle \langle k |+| k \rangle \langle \psii
|  \ \  {\rm or}  \ \ \hat{Y}_k = i | \psii \rangle \langle k | -
i\  | k \rangle \langle \psii| \ \,
\end{equation}
for all $k$ where  $\{| k \rangle \, |\, k=1, . . n-1 \} $ are a
set of vectors orthogonal to $|\psii \rangle$.

Similarly, we introduce a local chart on $SU(n)$ such that the
coordinates  are split  into a set of coordinates $\xi^\mu :
0,..n^2-2n$ for the isotropy subgroup and coordinates $y^\alpha$ ( $\alpha=1\, \ldots, 2(n-1)$ for the coset space $SU(n)/U(n-1)$. This we do
in order to  decompose  $\hatu \in SU(n)$ as
\begin{equation}\label{sundecomp}
\hatu = \hat{K}(y)\hat{H}(\xi)=
\hat{K}(y)\hat{H}_S(\xi_1,...\xi_{n^2-2n}) \, e^{i \gl_o \xi^o }
\end{equation}
where $H_s \in SU(n-1)$ and $\hat{K}(y)$ is a  coset
representative (for instance $\hat{K} = \exp[ i y^A \gl_A ]$). A section of states in Hilbert space is then
generated by the action of $\hat{K}$ on $\ket{\psi_1}$, i.e.,
\begin{equation}\label{raysection}
\ket{\psi(y) } \equiv \hat{K}(y)\ket{\psii} \, .
\end{equation}
Since $\hat{H}_s \ket{\psii} = \ket{\psii}$ and $\gl_o \ket{\psii}
= - \sqrt{ \frac{2(n-1)}{n}}\ket{\psii}$, it then follows from the
decomposition \rref{sundecomp} that  the amplitude $\mels{\hatu}$
may  be expressed in terms of the coset space amplitude
$\amp{\psif}{\psi(y) }$ according to
\begin{equation} \label{mateldecomp}
\mels{\hatu} = e^{- i\sqrt{ \frac{2(n-1)}{n}}\xi^0} \
\amp{\psif}{\psi(y) } \, .
\end{equation}

We proceed with the left invariant forms and the definition of the
Fubini-Study (FS) metric. For this we note that $\hat{H}^\dag
\D\hat{H}$ involves only the $\gl_i$, but $\hat{K}^\dag
\D\hat{K}$, not being a subgroup of $SU(n)$ expands as a linear
combination of all the group generators.  We therefore expand
$\hatu^\dag \D \hatu$ as
\begin{equation}
\hatu^\dag \D \hatu  \equiv  i \hat{H}^\dag\left[ \form{\tau}^A
\gl_A  + \form{a}^i \gl_i + \from^{i} \gl_i \right]\hat{H}\, ,
\end{equation}
where
\begin{eqnarray}
\form{\tau}^A & \equiv & \frac{1}{2 i}\tr{ \gl_A \hat{K}^\dag \D_{\pal} \hat{K}} \\
\form{\alpha}^i & \equiv &  \frac{1}{2 i} \tr{ \gl_i \hat{K}^\dag \D_{\pal} \hat{K}} \\
\from^i & \equiv &  \frac{1}{2 i}\tr{ \hat{H}^\dag \gl_i
\D_{\perp} \hat{H}}
\end{eqnarray}
and $\D_{\pal}$ and $\D_{\perp}$ denote external differentiation
with respect to the coset space ($y^A$) and subgroup ($\xi^i$)
coordinates respectively. Note that the vielbein $\from^i$ is defined in terms of  {\em right}-invariant forms and therefore differs from a corresponding left -invariant   form by a 
${\rm Ad}_H$ transformation which nonetheless leaves the $H$- Cartan-Killing  metric $ \eta_{ij}
\from^i \otimes \from^j $ invariant.

The Fubini-Study metric and its inverse are  defined by
considering the forms $\form{\tau}_A$ as vielbeins on the coset
space
\begin{equation}\label{fsdef}
\form{g}\sub{FS} = \eta\sub{AB} \form{\tau}^{A} \otimes
\form{\tau}^B \, , \ \ \ \form{g}^{-1}\sub{FS} = \eta^{AB}
\vect{t}_{A} \otimes \vect{t}_B
\end{equation}
where the vector fields $\vect{t}_A$ are dual to the
$\form{\tau}^A$, i.e., such that $\form{\tau}^A(\vect{t}_B) =
\delta^{A}_{B}$. Using \rref{raysection}, it is then a matter of some
algebra to show that the F.S. metric can be expressed as
 \begin{equation}
\form{g}\sub{FS} = \bra{\D \psi}\otimes\ket{\D \psi}_S - \amp{\D
\psi}{\psi}\otimes\amp{\psi}{\D \psi} \, .
\end{equation}
where $S$ stands for symmetrized. A similar calculation shows that since the Berry-Simon connection $\form{A}$ is $-i \amp{\psi_i}{\hat{K}^\dag
\D_{\pal} \hat{K}|\psi_i}$, it is related to $\form{\alpha}^0$ through
\begin{equation}\label{alphanot}
\form{\alpha}^0 =  \frac{1}{2 i} \tr{ \gl_0 \hat{K}^\dag
\D_{\pal} \hat{K}}=\sqrt{\frac{n}{2(n-1)}} \form{A} \, .
\end{equation}

Now, using $\form{g}\sub{CK} = \frac{1}{2}\tr{ \D U \otimes \D
U^\dagger }$, one can then show that the Cartan-Killing metric may
be written as
\begin{eqnarray}
\form{g}\sub{CK} & = & \form{g}\sub{FS} + \eta_{i j}
(\form{\alpha}^i + \from^i)\otimes(\form{\alpha}^i + \from^i) \, 
\end{eqnarray}
(note therefore that the  the $\vect{ t}_A$ are only orthonormal
with  respect to the FS metric). We shall also need the  inverse CK
metric, which is easily computed and is given by:
\begin{equation}
\form{g}\sub{CK}^{-1} = \eta^{AB} (\vect{t}_{A} - a_{A}^i \fre_j)
\otimes (\vect{t}_{A} - a_{A}^i \fre_j)  + \eta^{i j}
\fre_i\otimes \fre_j
\end{equation}
where $ a_{A}^i \equiv \form{\alpha}^i(\vect{t}_A) \, . $

Now, For any function $f(y,\xi)$ on the group manifold, we can
therefore write
\begin{equation}
g\sub{CK}^{-1}(\D f,\D f) = g\sub{FS}^{-1}\left(\DD_{\pal} f
,\DD_{\pal} f \right) + \eta^{ij}\nabla_i f \nabla_j f
\end{equation}
where
\begin{equation}
\DD_{\pal} f = \D_{\pal}f - \form{\alpha}^i \nabla_{i}f
\end{equation}
and where $\nabla_{i}f = \D_{\perp}f(\fre_i)$. In particular, we
look at the function
\begin{equation}
\mels{U}= \sqrt{p(y)}e^{i \phs(y,\xi_o)}
\end{equation}
from \rref{mateldecomp}, and where
$$ \phs(y,\xi^0) = \tilde{\phs}(y,\xi^0=0) -\sqrt{ \frac{2(n-1)}{n}} \xi^0\, .
$$
Noting that $\gl_o$ commutes with $\hat{H}_S$, we find that
\begin{equation}
\from^0  =\frac{1}{2 i}\tr{ \gl_i e^{-i \xi^0 \lambda^0} \D_{\perp} e^{i \xi^0 \lambda^0}} =  \D \xi^0  \, .
\end{equation}
Thus, it follows that $ \nabla_o \eta = \sqrt{\frac{2(n-1)}{n}} $
, and from \rref{alphanot} that
\begin{equation}
\DD_{\pal} \phs = \D_{\pal}\phs +
\sqrt{\frac{2(n-1)}{n}}\form{\alpha}^o
 = \D_{\pal}\phs +  \form{A} \, .
\end{equation}
Similarly, we find that $\DD_{\pal} p = \D_{\pal} p$. In this way,
relations \rref{cprel} can be obtained from \rref{surel} by
letting $\xi^o =0$ and using and
\begin{eqnarray}
g\sub{CK}^{-1}(\D \phs,\D \phs) & = & g\sub{FS}^{-1}(\DD_{\pal} \phs, \DD_{\pal} \phs)  + 2 \left(\frac{ n-1 }{n} \right) \\
g\sub{CK}^{-1}(\D \phs,\D p) & = &  g\sub{FS}^{-1}(\DD_{\pal} \phs,\D_{\pal} p)  \\
g\sub{CK}^{-1}(\D p,\D p)  & = &  g\sub{FS}^{-1}(\D_{\pal}p
,\D_{\pal} p)  \, ,
\end{eqnarray}
respectively. This corresponds to the choice $q=1$ in the definition of the FS metric.

\section{ Acknowledgements} The author wishes to thank Y. Aharonov, P. Mazur and the late J. Anandan for helpful discussions.

\end{document}